\documentstyle[12pt,aasms4]{article}

\def\Einstein{{\it Einstein}}
\def\Ginga{{\it Ginga}}
\def\ROSAT{{\it ROSAT}}
\def\ASCA{{\it ASCA}}
\def\ltsima{$\; \buildrel < \over \sim \;$}
\def\simlt{\lower.5ex\hbox{\ltsima}}            
\def\gtsima{$\; \buildrel > \over \sim \;$}
\def\simgt{\lower.5ex\hbox{\gtsima}}            

\begin{document}

\title{Discovery of a Large-Scale Abundance Gradient
in the Cluster of Galaxies AWM7 with ASCA}

\author{H.~{\sc Ezawa}\altaffilmark{1}, 
Y.~{\sc Fukazawa}\altaffilmark{1,2},
K.~{\sc Makishima}\altaffilmark{1,2}, 
T.~{\sc Ohashi}\altaffilmark{3},
F.~{\sc Takahara}\altaffilmark{3}, 
H.~{\sc Xu}\altaffilmark{1,4},
and N.~Y.~{\sc Yamasaki}\altaffilmark{3}}

\altaffiltext{1}{Department of Physics, University of Tokyo,
7-3-1 Hongo, Bunkyo-ku, Tokyo 113, Japan; ezawa@phys.s.u-tokyo.ac.jp}
\altaffiltext{2}{Research Center for the Early Universe (RESCEU),
University of Tokyo,
7-3-1 Hongo, Bunkyo-ku, Tokyo 113, Japan}
\altaffiltext{3}{Department of Physics, Tokyo Metropolitan University,
1-1 Minami-Ohsawa, Hachioji, Tokyo 192-03, Japan}
\altaffiltext{4}{Institute for Space and Astrophysics, Department of Physics,
Shanghai Jiao Tong University,
Huashan Road 1954, Shanghai 200030, PRC}

\begin{abstract}
A large-scale gradient in the metal abundance has been
detected with \ASCA\ from an X-ray bright cluster of galaxies AWM7. 
The metal abundance shows a peak of 0.5 solar at the center and
smoothly declines to $\simlt 0.2$ solar at a radius of 500 kpc.
The gas temperature is found to be constant at 3.8 keV.
The radial distribution of iron can be fit with a $\beta$-model with
$\beta \sim 0.8$ assuming the same core radius (115 kpc) as
that of the intracluster medium. The metal distribution in AWM7
suggests that the gas injected from galaxies is not efficiently mixed
in the cluster space and traces the distribution of galaxies. 
\end{abstract}


\section{Introduction}

Measurement of metal distribution in the hot intracluster medium (ICM
hereafter) is important in constraining the origin of metals, in
estimating how much mixing of the ICM has occurred, and in knowing the
precise amount of metals in the ICM.  The near-by X-ray bright cluster
AWM7 is a suitable object since its low temperature makes the
equivalent width of iron emission line high.  The redshift of the
central cD galaxy NGC1129 is 0.0176 (\cite{beers}).  X-ray emission
from this cluster has been studied with \Einstein\ 
(\cite{awm7-kriss}), \Ginga\ (\cite{tsuru-d}), \ROSAT\ 
(\cite{awm7-neumann}), and recently with \ASCA\ in the PV phase
(\cite{awm7-ohashi}; \cite{awm7-xu}).  The emission is described by a
thermal model with $kT = 3.8$ keV, excluding the center where the
\ROSAT\ PSPC data indicate a low temperature component.  The \ASCA\ 
data show that metal abundance is a factor of $\sim 1.5$ high at the
center ($r < 4^\prime$, \cite{awm7-xu}).  However, large-scale
distributions of temperature and abundance have not been studied with
sufficient sensitivity.

In this letter, we report on the results from multi-pointing
observations of AWM7 with \ASCA\ (\cite{asca}).  This is the first
example where a significant variation of metal abundance is found over
a scale of 500 kpc. An $H_0 = 50$ km s$^{-1}$ Mpc$^{-1}$ is employed
indicating 30.7 kpc for $1^\prime$ at AWM7, and a number fraction of
${\rm Fe}/{\rm H} = 4.68 \times 10^{-5}$ (\cite{angr}) is used for the
definition of the 1 solar iron abundance.

\section{Observations and results}

\ASCA\ observed AWM7 at 6 positions,
2 centered at 
$({\rm R.A.}, {\rm Dec.})_{\rm J2000}=
(2^{\rm h}54^{\rm m}59^{\rm s}, 41^\circ 37^{\rm m}40^{\rm s})$ and
$(2^{\rm h}53^{\rm m}24^{\rm s}, 41^\circ 36^{\rm m}34^{\rm s})$
on Aug. 7, 1993, during the PV-phase,
and 4 pointings centered at 
$({\rm R.A.}, {\rm Dec.})_{\rm J2000}=
(2^{\rm h}55^{\rm m}59^{\rm s}, 41^\circ 33^{\rm m}34^{\rm s})$,
$(2^{\rm h}52^{\rm m}12^{\rm s}, 41^\circ 33^{\rm m}26^{\rm s})$,
$(2^{\rm h}54^{\rm m}26^{\rm s}, 41^\circ 58^{\rm m}28^{\rm s})$, and 
$(2^{\rm h}54^{\rm m}27^{\rm s}, 41^\circ 23^{\rm m}25^{\rm s})$
on Feb. 10-11, 1994 during the AO-1 phase.
Each observation was performed with exposure time of $\sim 20\;{\rm ksec}$.

Xu et al. (1997) utilized the central pointing observation to
investigate the detailed properties of the central region of the
cluster.  In order to look into the hot gas properties in a much
larger scale, we analyzed the whole mapping data from 6 pointings by
\ASCA.  In this letter we concentrate on the data obtained with the
GIS (\cite{gis-ohashi}; \cite{gis-makishima}) which covers a larger
area than the SIS.  After the nominal event selection, an average of
20 blank sky data with point sources masked has been subtracted as the
background.  The systematic error in the generated GIS background flux
is estimated to be 4 \% rms (\cite{ikebe-d}; \cite{ikebe-ascanews}).

X-ray surface brightness distribution of AWM7 in the 0.7--10 keV range
measured with the GIS is clearly elongated along east-west direction
as shown by Neumann and B\"{o}hringer (1995), but no subclusters or
irregular patchiness is seen.  The data are fitted with the following
model. We first assume a $\beta$-model $S(r) = S_0 \left[ 1 + (r/a)^2
\right]^{-3\beta+1/2}$ where $r$ is the projected radius, $S_0$ is
the central surface brightness, and $a$ is the core radius,
respectively. This profile is then modified into an ellipse by either
expanding or shrinking by a factor of $\sqrt{0.8}$ in perpendicular
directions so that the minor and major axis ratio becomes 0.8 as given
by Neumann and B\"{o}hringer (1995).  Fitting the GIS data with this
model in the energy range 2--10 keV with a radius of 0 to $45^\prime$
from the center, we obtain the parameter values $a =
{3^\prime.75^{+0^\prime.78}_{-0^\prime.38}}$ and $\beta =
0.58^{+0.03}_{-0.02}$. The errors indicate the 90\% statistical errors
for single parameter.  These values are slightly different from those
obtained by \cite{awm7-neumann} ($a=3^\prime.32\pm 0^\prime.16$ and
$\beta = 0.53\pm 0.01$ with $3\sigma$ errors).  This is probably due
to the difference in the sensitive energy range and the field of view,
or maybe due to the accuracy of position determination of \ASCA\
(\cite{attitude}).  We will use the $\beta$-model parameters obtained
with the GIS data for the subsequent analysis.

\placefigure{fig:spectra} 

Spectral analysis has been carried out on data which are spatially
divided into concentric regions around the cluster center.  Figure
\ref{fig:spectra} shows the GIS pulse-height spectra for 5 annular
regions.  The strong Fe-K line in the central region confirms the
previous results (\cite{awm7-xu}).  Moreover, we can also see that the
Fe-K line equivalent width drops drastically towards the outer regions
beyond $r\sim 15^\prime$.  The result suggests a large scale abundance
gradient in $\sim 500{\;\rm kpc}$ order.  However, we need to examine
the stray-light and point-spread function (PSF) properties of the
X-ray telescope on \ASCA\ (XRT, \cite{xrt}) in detail to ensure the
result.

The spectrum obtained from each annular region is contaminated by the
photons arriving from different sky regions, even from outside of the
field of view. This is because the PSF of the XRT extends to more than
$10'$ and the effect of stray-light is significant (\cite{xrt}). In
the case of AWM7, approximately 30\% of photons accumulated in the
$r=10^\prime-15^\prime$ region originates in the central bright region
($r<10^\prime$).  To cope with this, we have carried out spectral fits
by calculating approximate response functions which take into account
the relative amount of the flux contamination between different
regions.  This is the way similar to the method applied to the \ASCA\
data of the Coma cluster (\cite{coma-honda}).  In the process of
response calculation, we assume a uniform energy spectrum with the
surface brightness profile obtained by the GIS data.  This response
compensates the flux contamination effect and enables us to derive the
true temperature and abundance if they are uniform over the entire
cluster.  Corrections will be made to the result if any deviation from
the uniformity is observed.

\placefigure{fig:temperature} 

We fit each spectrum from the concentric annular regions with a single
temperature Raymond-Smith model (\cite{rays}) with absorption.
Spectral parameters are listed in Table \ref{table:result}, and the
temperature profile is shown in Figure \ref{fig:temperature}.  The
temperature is constant at 3.8 keV out to $r = 40^\prime$ with a
fluctuation of only about 10\%. The value is consistent with the
previous \Ginga\ results (\cite{tsuru-d}). The central region is
somewhat cooler than the outer region as shown by Xu et al. (1997) and
Mushotzky et al. (1996), but not enough to contradict with the
isothermal assumption made above when the response function is
prepared.  The amount of $N_{\rm H}$ is consistent with the past
results (\cite{awm7-xu}; \cite{awm7-neumann}).  The gradient in the
equivalent width of Fe-K line shown in Figure \ref{fig:spectra} will
not affect the result on the temperature, since effects of the Fe-K
line equivalent width on the temperature are negligibly small compared
with those of the continuum.  The variation due to the systematic
error in background flux is estimated to be 10 \% for derived
temperature even for the outermost $25^\prime-40^\prime$ region.

As indicated in Figure \ref{fig:spectra}, the metal abundance is, on
the other hand, inconsistent with the ``uniform'' assumption employed
for the present response. The profile roughly agrees with that derived
by Xu et al. (1997). We therefore need to further correct the
equivalent width of the Fe-K line by evaluating the flux contamination
for the line component.  We estimate this effect with the ray-tracing
simulation for the XRT (\cite{raytr}). The GIS best-fit $\beta$-model
is employed for the intrinsic surface brightness distribution of the
continuum.  For an approximation, we only consider contamination from
inner regions and ignore that from outer regions.  Since Figure
\ref{fig:spectra} indicates an abundance gradient from the center to
outer regions, contamination at the Fe-K line energy would be more
dominated by that from inner regions. For each ring, inner regions
always contribute more than 70\% of the total contaminating photons.
The abundances in outer regions are lowered by this correction; in the
$60^\prime-80^\prime$ region, the abundance is reduced to about 60 \%
of the original value obtained by the spectral fit.  Calibration
uncertainty in the stray-light flux is approximately 20 \%
(\cite{ascanews-xrt}), giving a fractional error of typically about 5
\% in the resultant line or continuum flux.

\placetable{table:result} 
\placefigure{fig:abundance} 

The radial profile of the metal abundance
after the correction described above is shown in Table \ref{table:result}
and Figure \ref{fig:abundance}.  
The abundance reaches 0.5 solar in the innermost
$5^\prime$ region, which confirms the previous \ASCA\ results
(\cite{awm7-xu}; \cite{awm7-ohashi}), while the abundance drops to
half at $r\sim 15^\prime$ or at 460 kpc. The result is
highly significant since overall systematic error is less than 20 \%.

Finally, to look into the azimuthal dependence of the hot gas
parameters, we extracted spectra from 4 sectors with $90^\circ$ step
each in the azimuthal angles and the radius range of
$5^\prime-15^\prime$ or $15^\prime-40^\prime$. The inner ring shows no
azimuthal variation of temperature and abundance at a 90\% confidence
level. In the outer ring the temperature shows a peak-to-peak
variation of 1 keV around the mean of 3.8 keV with a 90\% error of 0.3
keV or 0.4 keV, and the abundance has too large an error to assess an
azimuthal variation.  Therefore, at least the azimuthal variation of
the ICM temperature is inferred to be less than $40$\%.

\section{Discussion}

The \ASCA\ observations of AWM7 have shown a radial drop of the metal
abundance over a scale of 500 kpc. The ICM temperature, on the other
hand, is found to be constant at 3.8 keV\@. The radial scale of the
abundance change is far larger than the size of the cD galaxy and
larger than the PSF of the XRT with a half power radius of $\sim
3^\prime$ (\cite{xrt}).  So far, significant change of abundance in
the central $< 100$ kpc in a cluster has been observed in Centaurus,
Virgo clusters and AWM7 itself (\cite{cen}; \cite{virgo};
\cite{awm7-xu}).  However, this is the first detection of a
significant metal segregation on the cluster scale of $\sim 500$ kpc.

In Figure \ref{fig:abundance}, we plotted the expected abundance when
the density distribution of metals follows the $\beta$-model with the
same core radius as the gas.  We find that $\beta \sim 0.8$ can
approximate the observed abundance gradient. This value is naturally
greater than that of the ICM, and close to that for galaxy
distribution derived for general clusters (\cite{beta-gal}).
Therefore, in AWM7 the metals seem to be tracing the galaxies.
Metzler \& Evrard (1994) also suggested the existence of the abundance
gradient caused by a steeper distribution of galaxies than the gas.
If metals are indeed injected from galaxies (\cite{arnaud};
\cite{tsuru-d}), this feature suggests that no strong mixing has
occurred in the ICM since the period of metal injection.

Let us briefly estimate how much iron can diffuse out in the Hubble
time after a point-like injection. We shall neglect the sedimentation
of iron based on the previous study by Rephaeli (1978).
Spitzer (1962) gives formula for diffusion constants of ions in a plasma.  
Since the mean deflection time of iron is estimated as
\begin{displaymath}
t_{D}= 
8\times 10^{12}\;{\rm sec} \times 
\left( \frac{n_{\rm p}}{{10^{-4}}\;{\rm cm}^{-3}} \right)^{-1} 
= 2.5 \times 10^5\;{\rm yr}, 
\end{displaymath}
the diffusion constant in a 4 keV plasma with a density of 10$^{-4}$
cm$^{-3}$ becomes about $10^{26}$ cm$^{2}$. This implies that iron can
diffuse in the ICM only out to 10 kpc at the maximum in the Hubble
time. It seems, therefore, natural that the metals stay where they are
injected and trace the distribution of galaxies, unless there is
no large-scale bulk motion in the ICM.
The data also suggests that the mixing due to galaxy motion is inefficient
(see also \cite{okazaki}).

Recent evidences of significant temperature variation in clusters of
galaxies suggest that these clusters have experienced mergers in the
past several Gyrs (e.g.  \cite{coma-honda}). Such a dynamical
interaction between clusters would cause large-scale mixing of the ICM
as well as temperature variation, as suggested from numerical merger
simulations (\cite{coma-ishizuka}).  The observed uniformity of the
ICM temperature and the large-scale gradient in the abundance, tracing
the galaxy distribution, supports the view that this cluster has been
at rest from the time of the metal injection.

No other clusters have been observed with enough sensitivity in their
outer regions. Bright near-by objects such as A1060
(\cite{a1060-tamura}) and Centaurs cluster (\cite{cen}) have been
studied only within 350 kpc from the center with \ASCA. Clusters
located at $z = 0.03 - 0.06$ are covered in a single GIS field, but
the data have typically an order of magnitude lower statistics than
the present case due to the strong vignetting of the XRT.  This
leads us to suspect that the large-scale abundance gradient could be
present in many other clusters. However, it is also possible that AWM7
is special because it has a very elongated morphology and probably has
a high ratio of gas to stellar mass (Neumann and B\"ohringer 1995).
Future systematic studies with \ASCA\ in the outer regions of other
near-by clusters would clarify whether the abundance gradient is
common or not.

This large-scale gradient in the metal distribution would have caused
an overestimation of the total mass of iron with the previous
non-imaging observations or with spatially non-slicing analysis.  The
measured abundance for AWM7, such as 0.42 with \Ginga\ 
(\cite{tsuru-d}: renormalized with Fe/H=4.68$\times$10$^{-5}$) and
0.39 with \ASCA\ excluding $r < 3'$ (\cite{awm7-mushotzky}), can be
significantly affected by the gradient.  If we approximate the metal
density distribution by a $\beta$-model with $\beta = 0.8$, the total
iron mass in the ICM within 1.2 Mpc from the center decreases by a
factor of 2.  An alternative model assuming a linear drop of abundance
from the center to 500 kpc following a constant level gives 0.18 solar
as the best-fit value with a 90\% error of $\pm 0.07$ solar for the
$15'-40'$ region. This model gives 10 \% increase of the iron mass to
the $\beta$-model estimation.  Tsuru (1992) shows that the total iron
mass in the ICM is proportional to the total stellar mass and that the
present supernova rate can account for only 10 \% of metals even
accumulating for the Hubble time.  If the metal distribution in the
ICM has a large-scale gradient as in AWM7 or a patchy nature in the
extreme case, the iron mass problem may be somewhat relaxed.  A fine
resolution (better than an arcminute) imaging of the metal
distribution in clusters is necessary to obtain the true amount of
metals in the cluster space.  Such a study would also reveal the
injection process and its history by the spatial size of the metal
distribution around galaxies or along the galaxy trajectories.

\acknowledgements

Special thanks are due to Dr. D.~M.~Neumann for discussion, and Dr.
H.~Honda, Dr. M.~Hirayama and Mr. K.~Kikuchi for valuable advices on
the analysis method.  HE acknowledges support by the Fellowships of
the Japan Society for Promotion of Science for Japanese Junior
Scientists.

\newpage

\begin{table}
\begin{center}
\small
\begin{tabular}{ccccccc}
\hline\hline
Radius  & $0-5^\prime$ & $5^\prime-10^\prime$ & $10^\prime-15^\prime$ &
   $15^\prime-20^\prime$ & $20^\prime-30^\prime$ & $25^\prime-40^\prime$ \\
\hline
Temperature [keV] &
$3.63\pm 0.09$ & $3.67^{+0.12}_{-0.09}$ & $3.83^{+0.15}_{-0.14}$ &
$3.74\pm 0.21$ & $4.10^{+0.32}_{-0.30}$ & $3.51^{+0.43}_{-0.37}$ \\
Abundance [solar] &
$0.51\pm 0.05$ & $0.38\pm 0.07$ & $0.30\pm 0.11$ &
$0.14 (<0.29)$ & $0.17 (<0.34)$ & $0.15 (<0.39)$ \\
$N_{\rm H}$ [$10^{20}\;{\rm cm}^{-2}$] &
$8.7\pm 1.7$ & $9.8^{+1.7}_{-1.8}$ & $8.3\pm 2.0$ &
$8.9^{+3.2}_{-2.5}$ & $8.0^{+3.5}_{-3.4}$ & $7.2^{+6.0}_{-5.4}$ \\
\hline
\end{tabular}
\end{center}
\caption{Results from the spectral fit for concentric annular regions
in AWM7. Response function is for an isothermal gas, which is a consistent
assumption with the listed results.
The abundances are corrected for the contamination due to PSF and 
stray-light as described in the text. 
The errors indicate 90\% statistical errors for a single parameter.
\label{table:result}}
\end{table}

\newpage

\figcaption[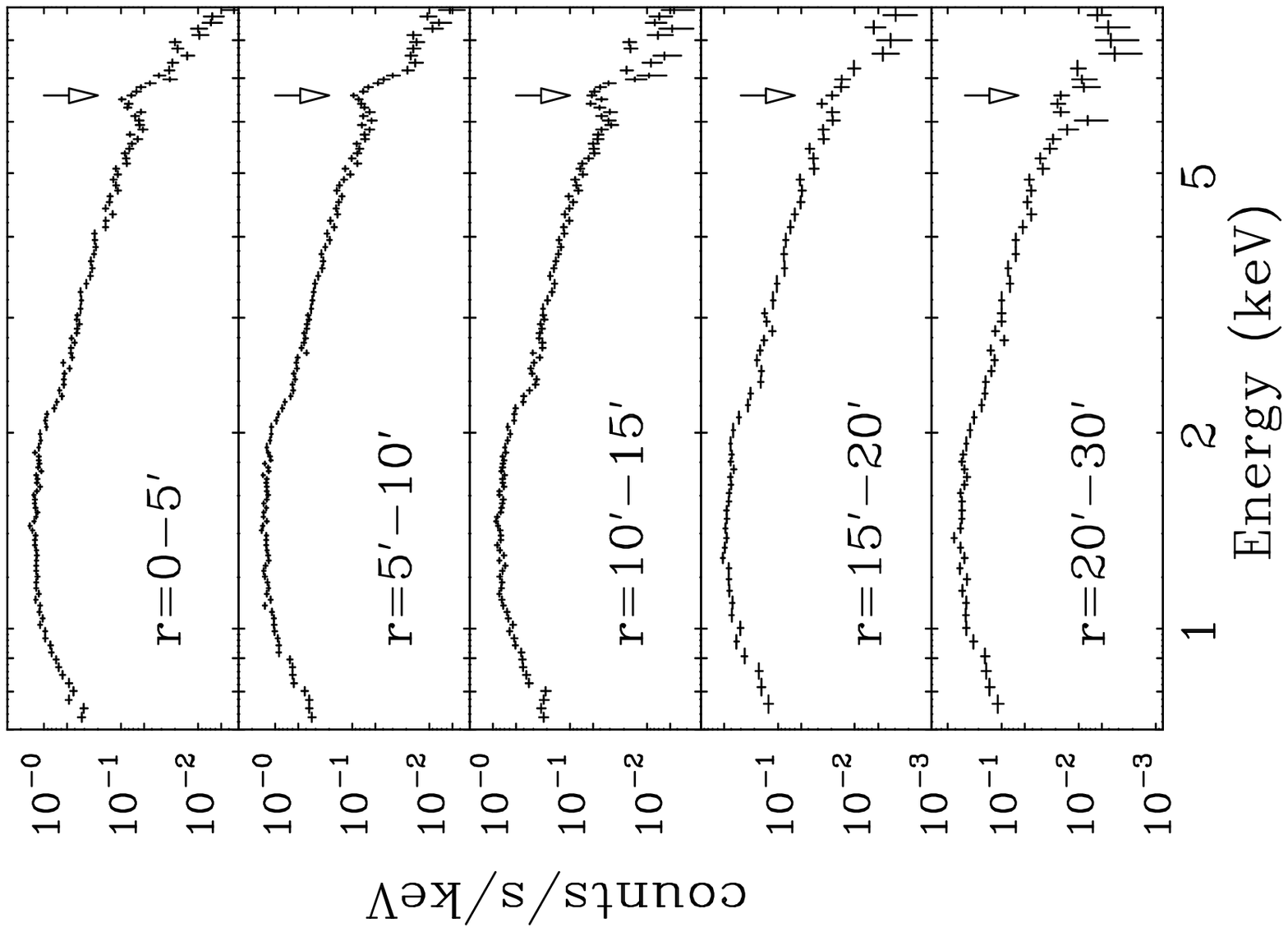]{Energy spectra obtained by the GIS from
5 concentric annular region in AWM7. The arrow in each spectrum
indicates the Fe-K emission line, which shows a gradual decrease of the
equivalent width with radius.
\label{fig:spectra}}

\figcaption[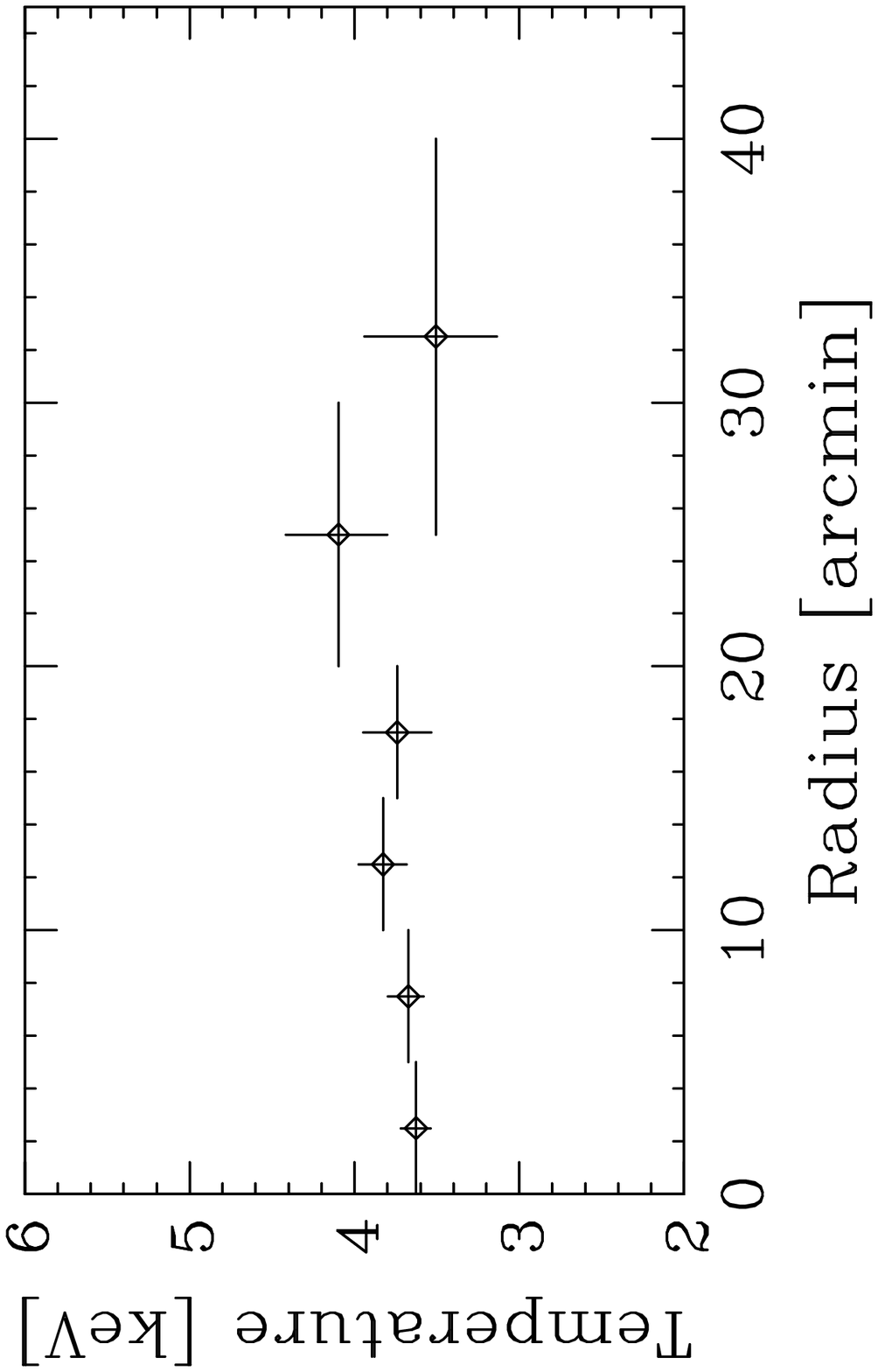]{Hot gas temperature in AWM7 for concentric 
annular regions.  Each spectrum is fitted with the response function
assuming a uniform spectrum as described in the text.
The error bars indicate 90 \% statistical errors.
\label{fig:temperature}}

\figcaption[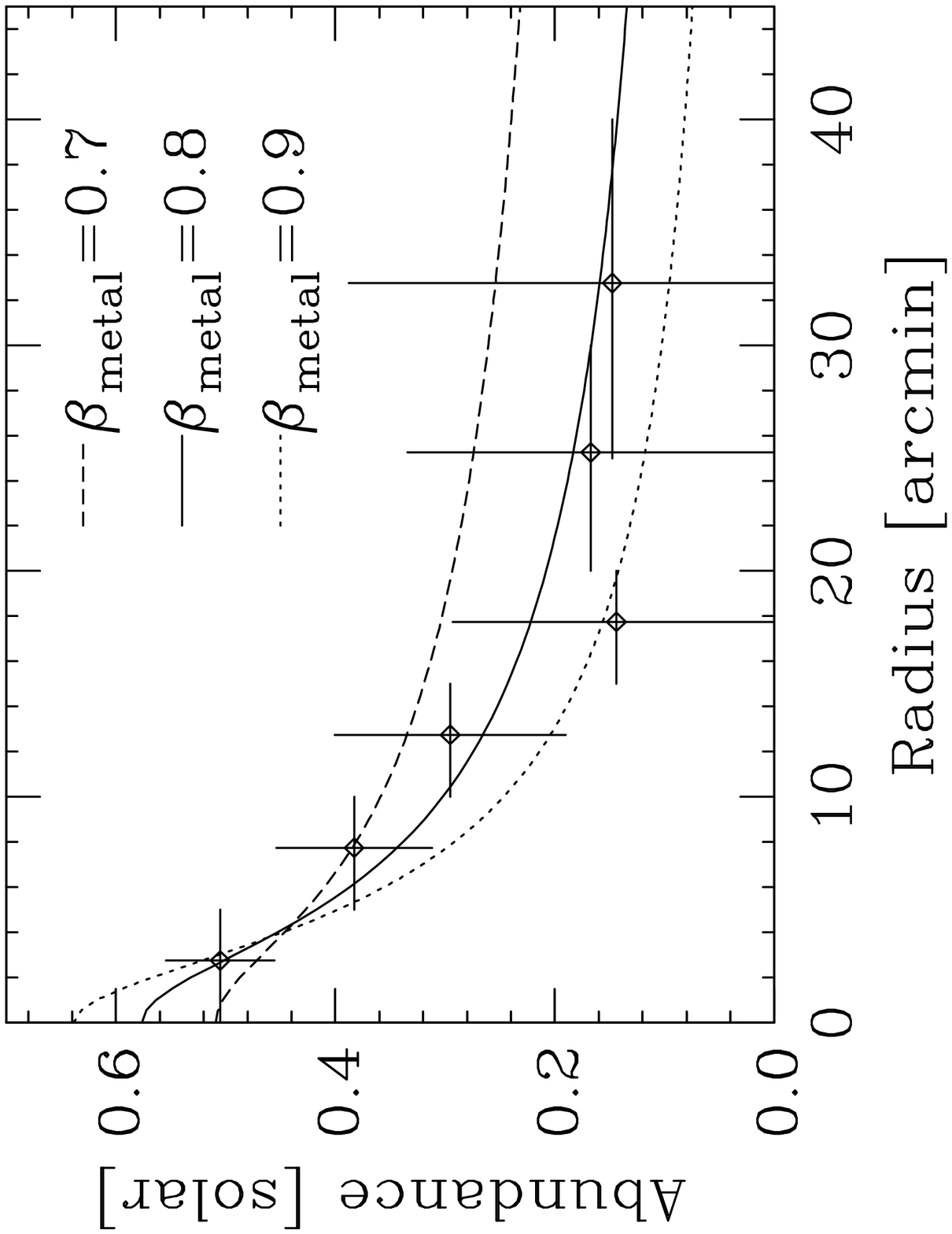]{Radial profile of the metal abundance in AWM7.
  The plotted values are corrected for the effects of PSF and stray
  light as described in the text.  The error bars are 90 \%
  statistical errors.  The curves indicate expected abundance profiles
  when mass densities of metals follow $\beta$-models modified into an
  ellipse having a minor to major axis ratio of 0.8, for 3 different
  values of $\beta_{\rm metal}$ with the same core radius with the
  gas. The observed abundance profile suggests that metal distribution
  follows $\beta_{\rm metal} \sim 0.8$.
\label{fig:abundance}}

\end{document}